\documentclass[12pt]{iopart}
\usepackage{citesort}

\usepackage{amssymb}
\usepackage{graphicx}
\usepackage{dcolumn}
\usepackage{bm}
\usepackage{color}
\usepackage{multirow}
\usepackage{graphicx}
\usepackage{placeins}
\usepackage{todonotes}
\begin{document}

\title[EIT of a single-photon in dipole-coupled 1D atomic clouds]{Electromagnetically induced transparency of a single-photon in dipole-coupled one-dimensional atomic clouds}
\author{Daniel Viscor, Weibin Li, and Igor Lesanovsky}
\address{School of Physics and Astronomy, The University of Nottingham, Nottingham, NG7 2RD, United Kingdom.}
\vspace{10pt}
\date{\today}

\begin{abstract}
We investigate the propagation of a single photon under conditions of electromagnetically induced transparency in two parallel one-dimensional atomic clouds which are coupled via Rydberg dipole-dipole interaction. 
Initially the system is prepared with a single delocalized Rydberg excitation shared between the two ensembles and the photon enters both of them in an arbitrary path-superposition state. 
By properly aligning the transition dipoles of the atoms of each cloud we show that the photon can be partially transferred from one cloud to the other via the dipole-dipole interaction. 
This coupling leads to the formation of dark and bright superpositions of the light which experience different absorption and dispersion. 
We show that this feature can be exploited to filter the incident photon in such a way that only a desired path-superposition state is transmitted transparently. 
Finally, we generalize the analysis to the case of $\mathcal{N}$ coupled one-dimensional clouds. We show that within some approximations the dynamics of the propagating photon can be mapped on that of a free particle with complex mass.
\end{abstract}


\section{Introduction}
\label{sec:Introduction}

Propagation of light in ensembles of Rydberg atoms \cite{saffman_quantum_2010}, i.e., atoms in higly-excited states, under conditions of electromagnetically induced transparency (EIT) \cite{fleischhauer_EIT_2005} has been widely investigated in recent years \cite{gorshkov_dissipative_2013,pritchard_cooperative_2010,peyronel_2012,Tiarks_transistor_2014,Gorniaczyk_transistor_2014,baur_single_2014,he_two-photon_2014,Bienias_Scattering_2014}. The intense activity focused on this subject has led to a wide variety of important applications for single-photon non-linear optics and for optical quantum information processing. For instance, single-photon filters \cite{gorshkov_dissipative_2013,peyronel_2012} and substractors \cite{Honer_substractor_2011,gorshkov_photon-photon_2011}, photon transistors \cite{Tiarks_transistor_2014,Gorniaczyk_transistor_2014,Chen_2013}, switches \cite{baur_single_2014,Chen_2013}, and photon gates \cite{friedler_deterministic_2005,gorshkov_photon-photon_2011,ParedesBarato_AllOptical_2014} have been implemented. 
All these applications are based on the the strong and long-ranged interaction between atoms in Rydberg states, which translates into an effective photon-photon interaction \cite{firstenberg_attractive_2013,gorshkov_photon-photon_2011,he_two-photon_2014}.
In general, most of these effects arise from a density-density interaction between Rydberg states. However, increasing interest is currently devoted to the study of the resonant dipole-dipole interaction (DDI) between different Rydberg states \cite{walker_consequences_2008}, the so-called F\"orster resonances \cite{nipper_highly_2012,ryabtsev_forster_2010}. The resulting strong interactions have been exploited recently, for instance, to implement efficient photon transistors \cite{Tiarks_transistor_2014,Gorniaczyk_transistor_2014}. Beyond that, the  coupling between different Rydberg states forms the basis of coherent population exchange which has recently been studied by a number of groups \cite{anderson_resonant_1998,Ditzhuijzen08,mudrich_transfer_2010,gunter_observing_2013,ravets_2014}. Moreover, it was shown theoretically that the corresponding exchange interaction can generate an enhancement of the optical depth and leads to non-local light propagation effects in Rydberg EIT \cite{Entangled_EIT_2014}.

\begin{figure}
\centering
\includegraphics*[width=0.8\columnwidth]{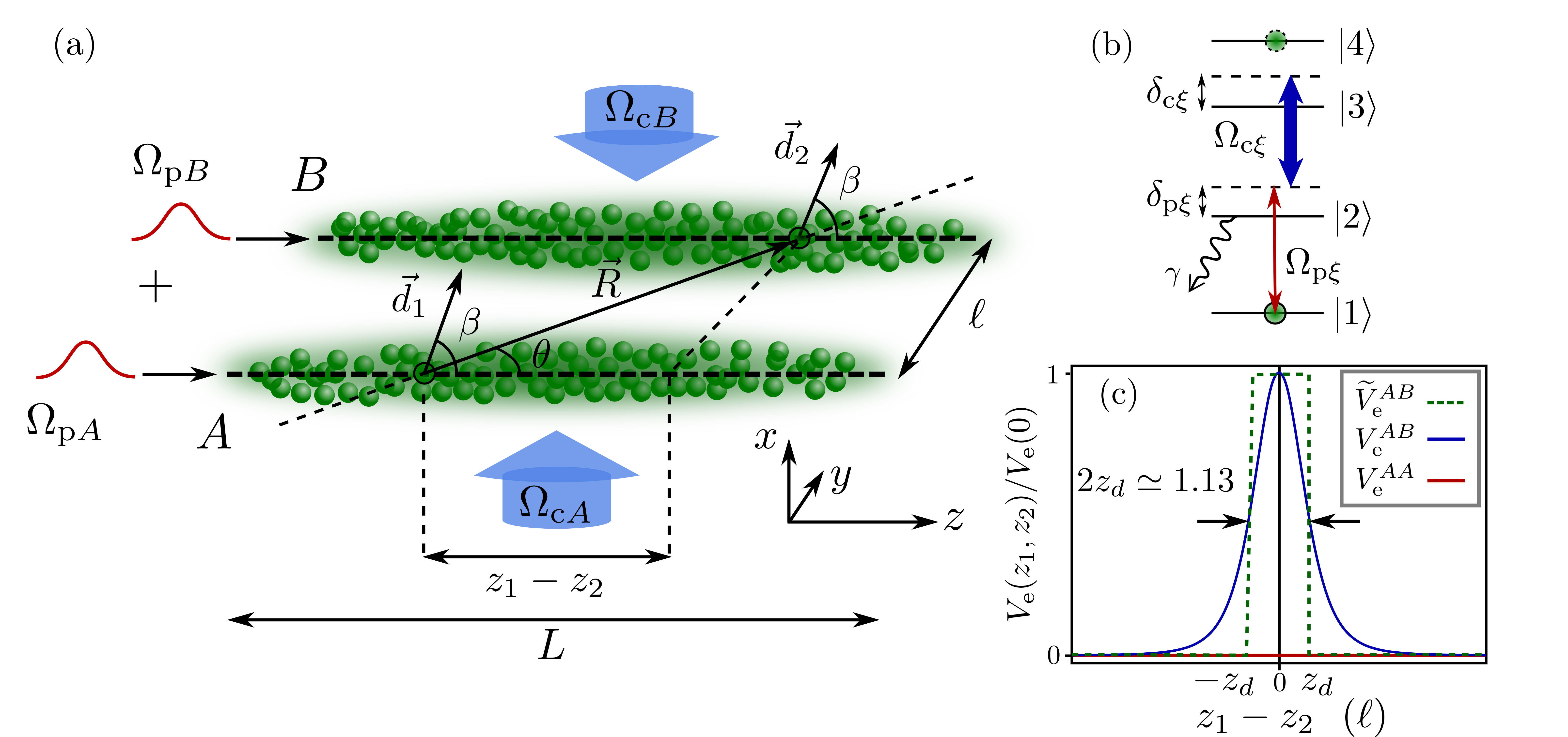}
\caption{(a) Single photon entering an atomic medium --- consisting of two  clouds --- in a path-superposition state (PSS). The two clouds ($A$ and $B$) are located in the $y$--$z$ plane and are separated by a distance $\ell$. The atomic transition dipoles $\vec{d}_{1}$ and $\vec{d}_{2}$, placed at positions $z_1$ and $z_2$, respectively, are aligned along the quantization axis $\vec{e}_q$ and separated by the vector $\vec{R}$. The quantization axis $\vec{e}_q$ is chosen to lie in the $x$--$z$ plane, forming an angle $\beta$ with the propagation axis $+z$. (b) The atoms of the medium have four relevant levels in a ladder configuration. An initial single excitation which is coherently shared among the two clouds, is prepared in state $|4\rangle$. The remaining states, $|1\rangle$, $|2\rangle$ and $|3\rangle$, form an EIT configuration: The transition $|1\rangle\leftrightarrow|2\rangle$ ($|2\rangle\leftrightarrow|3\rangle$) is coupled by the single-photon (strong-control) field of Rabi frequency $\Omega_{\rm p\xi}$ ($\Omega_{\rm c\xi}$), with detuning $\delta_{\rm p\xi}$ ($\delta_{\rm c\xi}$). State $\left|2\right\rangle$ is spontaneously decaying at a rate $\gamma$. The DDI couples atomic pair states, e.g. $|3\rangle_i|4\rangle_j\leftrightarrow|4\rangle_i|3\rangle_j$. (c) DDI strength for atoms in the same (red line) and different (blue line) clouds, as a function of the axial offset $z_1-z_2$ for $\beta=\arccos(1/\sqrt{3})$. The dashed line represents an approximation to the inter-cloud DDI $\widetilde{V}^{AB}_{\mathrm{e}}$ (see details in  Sec. \ref{sec:ApproximateAnalyticalSolution}). Both the actual and the approximate DDI strengths have the same full width at half maximum $2z_d\simeq1.13\ell$.}
\label{fig:fig1}
\end{figure}

In this paper, connecting to the recent work \cite{Entangled_EIT_2014}, we study a system of two parallel atomic clouds that are prepared in a spinwave state \cite{dudin_emergence_2012,dudin_strongly_2012}. Within this setup we investigate the propagation, under EIT conditions, of a single photon entering both ensembles simultaneously as shown in Fig.~\ref{fig:fig1}(a).
In contrast with other similar setups \cite{wilk_entanglement_2010,Isenhower_CNOT_2010,he_two-photon_2014,ParedesBarato_AllOptical_2014,Wu_EIT_VdWinteracion_2014}, where atoms or groups of spatially separated atoms interact via a density-density interaction, we exploit here features of the angular dependence of the resonant DDI between different atomic angular-momentum states \cite{Carroll_angular_2004,reinhard_level_2007,saffman_efficient_2009,saffman_quantum_2010,Barredo_3atom_anisotropic_2014}. 
Specifically, we envision a situation in which the atoms in one cloud can only interact with the atoms in the opposite one via resonant DDI, which leads to a coherent atomic population exchange between the two clouds. The population exchange is one of the key ingredients of our setup because it leads to exchange among the two propagating components of the incident photon field. This in turn results in vastly different optical properties for the symmetric and antisymmetric path superposition of the propagating single photon. We will show that by properly choosing the system parameters, it is possible to transmit only a given superposition state (e.g., the antisymmetric component) thus realizing a tunable single-photon path-superposition state (PSS) filter. Finally, we generalize the previous system to $\mathcal{N}$ parallel clouds. Here we find that the dynamics of the photon resembles that of a free particle with complex mass.

The paper is organized as follows. First, in Sec.~\ref{sec:PhysicalSystem} we introduce the physical system under consideration and discuss the Hamiltonian that determines its evolution. Second, in Sec.~\ref{sec:SinglePhotonFiltering} we derive the evolution equations for the atomic amplitudes and the light field propagating in the two clouds. We obtain the susceptibility of the medium, and derive an approximate analytical solution for the light propagation in terms of symmetric and antisymmetric normal modes. We find that both modes experience different absorption and acquire different group velocities. This is confirmed by numerical calculations that show good agreement with the analytical predictions. In Sec.~\ref{sec:GeneralizationToNClouds} we discuss the generalization of the setup to $\mathcal{N}$ parallel clouds. Finally, in Sec.~\ref{sec:SummaryAndConlusions} we summarize the results and present the conclusions.

\section{Physical system}
\label{sec:PhysicalSystem}

\subsection{Atomic level structure and initial condition}
The system under consideration is depicted in Fig.~\ref{fig:fig1}(a). It consists of two parallel one-dimensional (1D) atomic clouds, $A$ and $B$, of length $L$ which are located in the $y$--$z$ plane and separated by a distance $\ell$. The atoms have four relevant energy levels in a ladder configuration, as shown in Fig.~\ref{fig:fig1}(b). The lower levels $\left|1\right\rangle$ and $\left|2\right\rangle$ are, respectively, a ground state and an excited low-lying state, where the latter has a spontaneous decay rate $\gamma$. The two uppermost levels $\left|3\right\rangle$ and $\left|4\right\rangle$ correspond to long-lived $nS$ and $n'P$ states, respectively, with high principal quantum numbers $n, n'$. We assume that initially the entire atomic medium is prepared in a ``superatom'' state of the form 
\begin{eqnarray}
	\left|\psi_{1}(t=0)\right\rangle &=& \sum_{j}c^{(j_A)}_{4}(0)\left|1,...,4_j,...,1\right\rangle_A\left|1,...,1\right\rangle_B \nonumber \\
	&&+ \sum_{j}c^{(j_B)}_{4}(0)\left|1,...,1\right\rangle_A\left|1,...,4_j,...,1\right\rangle_B.
\label{eq:SpinwaveState}
\end{eqnarray}
This state describes a collectively shared single excitation \cite{dudin_strongly_2012,dudin_emergence_2012} in state $\left|4\right\rangle$ with the coefficients $c^{j_A}_{4}(0)$ and $c^{j_B}_{4}(0)$ being the probability amplitudes of finding the excitation in an atom $j$ of cloud $A$ or $B$, respectively. Such state can be prepared, for instance, by means of the Rydberg blockade \cite{saffman_quantum_2010} or by using attenuated laser pulses.

With this initial condition we consider a single-photon entering the two clouds along the $+z$ direction, which is initially in an arbitrary superposition state of two wavepackets propagating along different clouds, i.e., a PSS. The photon couples the $\left|1\right\rangle\leftrightarrow\left|2\right\rangle$ transition of atoms in cloud $A$ $(B)$ with a detuning $\delta_{{\rm p}A}$ ($\delta_{{\rm p}B}$). The adjacent transition $\left|2\right\rangle\leftrightarrow\left|3\right\rangle$ is driven by a strong control field of Rabi frequency $\Omega_{{\rm c}A}$ ($\Omega_{{\rm c}B}$) with a detuning $\delta_{{\rm c}A}$ ($\delta_{{\rm c}B}$), thus forming an EIT configuration. Note that with this setup the system can never contain more than two Rydberg excited atoms (the initial excitation in $\left|4\right\rangle$ and the one created by the single photon, in state $\left|3\right\rangle$) at the same time.

\subsection{Atomic interaction potential}
In this work, we consider a particular situation where the atoms in one cloud cannot interact with each other, but only with other atoms in the opposite cloud. This is achieved by exploiting the angular dependence of the DDI \cite{saffman_quantum_2010,Carroll_angular_2004,reinhard_level_2007,saffman_efficient_2009,Barredo_3atom_anisotropic_2014}. For two atoms 1 and 2 with transition dipole moments $\vec{d}_1$ and $\vec{d}_2$, respectively, separated by the relative vector $\vec{R}$, the DDI has the form:
\begin{eqnarray}
	\hat{V}_{\mathrm{dd}} & = & \frac{1}{4\pi\epsilon_0}\left[\frac{\vec{d}_{1}\vec{d}_{2}}{R^{3}}-3\frac{(\vec{d}_{1}\vec{R})(\vec{d}_{2}\vec{R})}{R^{5}}\right].
\label{eq:Dipole-dipole_int}
\end{eqnarray}
Here $R\equiv|\vec{R}|$, and $\epsilon_0$ is the electric permittivity in vacuum. To be specific we consider the configuration of Fig.~\ref{fig:fig1}(a). We choose states $\left|3\right\rangle\equiv |nS_{1/2}\rangle$ and $\left|4\right\rangle\equiv |n'P_{1/2}\rangle$, whose transition is driven by linearly polarized photons, and the control field being polarized along the quantization axis $\vec{e}_q$. We also consider that $\vec{e}_q$ is located in the $x$--$z$  plane at an angle $\beta$ with respect to the propagation axis $+z$ [see Fig.~\ref{fig:fig1}(a)].
With this choice only the non-diagonal matrix elements of the DDI contribute to the interaction as the contributions to $\hat{V}_{\mathrm{dd}}$ from components orthogonal to $\vec{e}_q$ vanish due to selection rules. 
This implies that the DDI only couples states $\left|3\right\rangle$ and $\left|4\right\rangle$. Therefore, for the setup given in Fig.~\ref{fig:fig1}(a), its operator form is
\begin{eqnarray}
	\hat{V}_{\mathrm{dd}}&=&
	\hbar V^{AB}_{\mathrm{e}}(z_1,z_2)\left|4\right\rangle_{1_A}\left\langle 3\right|\otimes\left|3\right\rangle_{2_B}\left\langle 4\right|+
	\hbar V^{AA}_{\mathrm{e}}(z_1,z_2)\left|4\right\rangle_{1_A}\left\langle 3\right|\otimes\left|3\right\rangle_{2_A}\left\langle 4\right|\nonumber \\
	&&+\hbar V^{BB}_{\mathrm{e}}(z_1,z_2)\left|4\right\rangle_{1_B}\left\langle 3\right|\otimes\left|3\right\rangle_{2_B}\left\langle 4\right|+{\rm h.c.}, 
\label{eq:Vdd}
\end{eqnarray}
where
\begin{equation}
	V^{\mu\nu}_{\mathrm{e}}(z_1,z_2)=\frac{C_3}{\left[(z_1-z_2)^2+\ell_{\mu\nu}^2\right]^{3/2}}\left[1-3\cos^2\left(\beta\right)\frac{(z_1-z_2)^2}{(z_1-z_2)^2+\ell_{\mu\nu}^2}\right]
\label{eq:Ve}
\end{equation}
is the interaction strength, $C_3=\langle\vec{d}_{1}\vec{d}_{2}\rangle/(4\pi\epsilon_0\hbar)$ is the so called dispersion coefficient, and $\ell_{\mu\nu}=\ell(1-\delta_{\mu\nu})$ with $\delta_{\mu\nu}$ being the Kronecker delta function. To finally achieve the situation where only atoms located in different clouds are interacting, we choose $\beta=\arccos\left(1/\sqrt{3}\right)$. As shown in Fig.~\ref{fig:fig1}(c) (red curve) the interaction among atoms of the same cloud is indeed completely suppressed (notice that we have not considered second-order dipole-dipole ($1/R^6$) interactions that would lead to van der Waals interactions between atoms in the same cloud). 
On the other hand, for atoms belonging to different clouds the interaction exhibits a symmetric peak profile as a function of $\Delta z$ (blue curve) whose full width at half maximum (FWHM) is given by $2z_d$, with $z_d=\ell\sqrt{4^{1/5}-1}$.

\subsection{Atomic Hamiltonian}
With these considerations the Hamiltonian of the system in the slowly varying envelope and rotating-wave approximations becomes:
\begin{eqnarray}
	\hat{H} & = & \sum_{\mu=A,B}(\hat{H}^{(\mu)}_{0}+\hat{H}^{(\mu)}_{I})+\hat{H}_{D} \label{eq:TotalH}.
\end{eqnarray}
Here
\begin{eqnarray}
	\hat{H}_{0}^{(\mu)} & = & \sum_{j=1}^{N_{\mu}}\hbar\left[\omega_{21}\hat{\sigma}_{22}^{\left(j_{\mu}\right)}+\omega_{31}\hat{\sigma}_{33}^{\left(j_{\mu}\right)}+\omega_{41}\hat{\sigma}_{44}^{\left(j_{\mu}\right)}\right],	\label{eq:AtomicH}\\
	\hat{H}_{I}^{(\mu)} & = & -\sum_{j=1}^{N_{\mu}}\hbar\left[\Omega_{\rm p\mu}e^{-i\omega_{\rm p\mu}^{0}t}\hat{\sigma}_{21}^{\left(j_{\mu}\right)}+ \Omega_{\rm c\mu}e^{-i\omega_{\rm c\mu}^{0}t}\hat{\sigma}_{32}^{\left(j_{\mu}\right)}+\mathrm{h.c.}\right], \\
	\hat{H}_{D} & = & 
	\sum_{i=1}^{N_{A}}\sum_{j=1}^{N_{B}}\hbar V^{AB}_{\rm e}\left(z_i,z_j\right)\left[\hat{\sigma}_{34}^{\left(i_{A}\right)}\hat{\sigma}_{43}^{\left(j_{B}\right)}+\mathrm{h.c.}\right]	\label{eq:RydIntH}
\end{eqnarray}
and $N_{\mu}$ is the number of atoms in cloud $\mu$. Moreover, we have defined $\omega_{ij}\equiv\omega_i-\omega_j$, with $\omega_j$ being the frequency of state $\left|j\right\rangle$ ($\hbar\omega_{1}$ is taken as the energy reference) and $\hat{\sigma}_{lm}^{\left(j_\mu\right)}\equiv\left|l\right\rangle_{j_\mu}\left\langle m\right|_{j_\mu}$ being the transition operator for atom $j$, placed at position $z_{j}$ of cloud $\mu$.
The Rabi frequencies for the probe (control) field are defined as $\Omega_{\rm p\mu}\equiv\vec{d}_{12}\vec{E}_{\rm p\mu}/(2\hbar)$ [$\Omega_{\rm c\mu}\equiv\vec{d}_{23}\vec{E}_{\rm c\mu}/(2\hbar)$], with $\vec{d}_{12}$ ($\vec{d}_{23}$) being the dipole moment of the $\left|1\right\rangle\leftrightarrow\left|2\right\rangle$ ($\left|2\right\rangle\leftrightarrow\left|3\right\rangle$) transition, and $\vec{E}_{\rm p\mu}$ ($\vec{E}_{\rm c\mu}$) being the probe (control) electric field, which oscillates at a central frequency $\omega^0_{\rm p\mu}$ ($\omega^0_{\rm c\mu}$). The spatial and temporal dependencies of the control and probe fields will be considered in the next section.

\section{Single-photon dynamics and PSS filtering}
\label{sec:SinglePhotonFiltering}

In this section we explore the dynamics of the system and demonstrate its potential application as a single-photon PSS filter. We employ a semiclassical treatment based on the Schr\"odinger and Maxwell equations. In this model, the radiative decay rate $\gamma$ from state $\left|2\right\rangle$ is treated perturbatively and included in the model by replacing the detuning $\delta_{\rm p\mu}$ by the complex quantity $\Delta_{\rm p\mu}=\delta_{\rm p\mu}+i\gamma$ \cite{Linington_ComplexDetuning_2008}.
This approach allows us to work in terms of the state vector probability amplitudes instead of the density matrix elements. This is, moreover, equivalent to considering that spontaneous emission removes population from the four-level system. This leads to a reduction of the norm of the state vector which is interpreted as absorption. Note that due to the form of the initial state Eq.~(\ref{eq:SpinwaveState}) and the typically justified assumption that the lifetime of the high-lying states is longer than the pulse propagation time inside the medium, there will be always one atom populated in state $|4\rangle$.

\subsection{Evolution equations}
\label{sec:EvolutionEquations}
To obtain the equations of motion of the atomic probability amplitudes we use the fact that the quantum state of the atoms at any time is given by the wavefunction $\left|\psi\left(t\right)\right\rangle = \left|\psi_{1}\left(t\right)\right\rangle + \left|\psi_{2}\left(t\right)\right\rangle + \left|\psi_{3}\left(t\right)\right\rangle$ (a similar approach is for instance pursued in \cite{Moiseev_QuantumMemory_2004}), where the terms at the right hand side have the form
\begin{eqnarray}
	\left|\psi_{1}(t)\right\rangle  =&  \sum_{i}\left[c_{4}^{\left(i_{A}\right)}(t)\hat{\sigma}_{41}^{\left(i_{A}\right)} +  c_{4}^{\left(i_{B}\right)}(t)\hat{\sigma}_{41}^{\left(i_{B}\right)}\right]\left|\bar{1}\right\rangle, \label{eq:psi1} \\
	\left|\psi_{2}(t)\right\rangle  =&  \sum_{i,j}\left[c_{24}^{\left(i_{A},j_{B}\right)}(t)\hat{\sigma}_{21}^{\left(i_{A}\right)}\hat{\sigma}_{41}^{\left(j_{B}\right)} + c_{42}^{\left(i_{A},j_{B}\right)}(t)\hat{\sigma}_{41}^{\left(i_{A}\right)}\hat{\sigma}_{21}^{\left(j_{B}\right)}\right]\left|\bar{1}\right\rangle \nonumber \\
	&  +\sum_{i,j\neq i}\left[c_{42}^{\left(i_{A},j_{A}\right)}(t)\hat{\sigma}_{41}^{\left(i_{A}\right)}\hat{\sigma}_{21}^{\left(j_{A}\right)} + c_{24}^{\left(i_{B},j_{B}\right)}(t)\hat{\sigma}_{21}^{\left(i_{B}\right)}\hat{\sigma}_{41}^{\left(j_{B}\right)}\right]\left|\bar{1}\right\rangle, \label{eq:psi2} \\
	\left|\psi_{3}(t)\right\rangle  =&  \sum_{i,j}\left[c_{34}^{\left(i_{A},j_{B}\right)}(t)\hat{\sigma}_{31}^{\left(i_{A}\right)}\hat{\sigma}_{41}^{\left(j_{B}\right)} + c_{43}^{\left(i_{A},j_{B}\right)}(t)\hat{\sigma}_{41}^{\left(i_{A}\right)}\hat{\sigma}_{31}^{\left(j_{B}\right)}\right]\left|\bar{1}\right\rangle \nonumber \\
	&  +\sum_{i,j\neq i}\left[c_{43}^{\left(i_{A},j_{A}\right)}(t)\hat{\sigma}_{41}^{\left(i_{A}\right)}\hat{\sigma}_{31}^{\left(j_{A}\right)} + c_{34}^{\left(i_{B},j_{B}\right)}(t)\hat{\sigma}_{31}^{\left(i_{B}\right)}\hat{\sigma}_{41}^{\left(j_{B}\right)}\right]\left|\bar{1}\right\rangle, \label{eq:psi3}
\end{eqnarray}
with $\left|\bar{1}\right\rangle\equiv\bigotimes_j\left|1\right\rangle_{j_A}\left|1\right\rangle_{j_B}$. The sums are taken over all the atoms and the coefficients $c_{lm}^{(i_\mu,j_{\nu})}(t)$ are the many-particle probability amplitudes of the different states.
The first term [Eq.~(\ref{eq:psi1})] corresponds to the system having one atom in state $|4\rangle$ and the rest of the atoms being in the ground state $|1\rangle$. The second [Eq.~(\ref{eq:psi2})] and third [Eq.~(\ref{eq:psi3})] terms correspond to the same situation but with an additional excitation in states $|2\rangle$ and $|3\rangle$, respectively. 

By inserting Eqs.~(\ref{eq:psi1})-(\ref{eq:psi3}) into the Schr\"odinger equation one obtains the evolution equations for the probability amplitudes. Next we change to a frame which rotates at frequency $(\omega_{41}+\omega_{\rm p\mu}^{0})$ for $\left|\psi_2\right\rangle$ and at $(\omega_{41}+\omega_{\rm p\mu}^{0}+\omega_{\rm c\mu}^{0})$ for $\left|\psi_3\right\rangle$. Furthermore, by considering a sufficiently dense atomic gas, we employ a spatially continuous description by introducing continuous versions of the probability amplitudes
\begin{eqnarray*}
  c^{(i_{\mu},j_{\nu})}_{24}(t)\rho_{\mu}(z)\rho_{\nu}(z')e^{i\omega_{41}t}e^{i\omega_{\rm p\mu}^{0}t}&\rightarrow&\alpha^{\mu\nu}_{24}\left(t,z,z'\right)\\
  c^{(i_{\mu},j_{\nu})}_{34}(t)\rho_{\mu}(z)\rho_{\nu}(z')e^{i\omega_{41}t}e^{i(\omega_{\rm p\mu}^{0}+\omega_{\rm c\mu}^{0})t}&\rightarrow&\alpha^{\mu\nu}_{34}\left(t,z,z'\right).
\end{eqnarray*}
These encode the probability of an atom at position $z$ of cloud $\mu$ being in state $\left|2\right\rangle$ and $\left|3\right\rangle$, respectively, and another at $z'$ of cloud $\nu$ being in $\left|4\right\rangle$ ($\mu,\nu=\left\{A,B\right\}$), with the remaining atoms being in the ground state. Here $\rho_{\mu,\nu}(z)$ is the linear single-particle density in cloud $\mu,\nu$. Accordingly, the probability amplitude of an atom being in state $\left|4\right\rangle$ is replaced by $c^{(i_{\mu})}_{4}(t)\rho_{\mu}(z)e^{i\omega_{41}t}\rightarrow\alpha^{\mu}_{4}\left(z,t\right)$. Finally, to simplify the equations we consider only terms up to first order in the probe field Rabi frequency $\Omega_{\rm p\mu}$ \cite{fleischhauer_dark-state_2000}. This implies that $\partial_{t}\alpha_{4}^{\mu}=0$, i.e., the initial excitation probability amplitude of atoms being in state $|4\rangle$ is a constant in each could, $\alpha_{4}^{\mu}(z,t)=\alpha_{4}^{\mu}(z)$.
After these manipulations the evolution equations for the probability amplitudes read
\begin{eqnarray}
	\partial_{t}\alpha^{\mu\nu}_{24}\left(z,z',t\right) = & i\Delta_{\rm p\mu}\alpha^{\mu\nu}_{24}\left(z,z',t\right)+i\Omega_{\rm p\mu}\left(z,t\right)\alpha^{\nu}_{4}\left(z'\right)  \nonumber \\ 
	&+i\Omega_{\rm c\mu}^{*}(z)\alpha^{\mu\nu}_{34}\left(z,z',t\right), \label{eq:alpha24AB} \\
	\partial_{t}\alpha^{\mu\nu}_{34}\left(z,z',t\right) = & i\Delta_{\rm c\mu}\alpha^{\mu\nu}_{34}\left(z,z',t\right)+i\Omega_{\rm c\mu}(z)\alpha^{\mu\nu}_{24}\left(z,z',t\right) \nonumber \\
	&-iV^{\mu\nu}_{\rm e}\left(z,z'\right)\alpha^{\nu\mu}_{34}\left(z',z,t\right). \label{eq:alpha34AB}
\end{eqnarray}
Here we have defined the detunings $\Delta_{\rm c\mu}=\delta_{\rm p\mu}+\delta_{\rm c\mu}$, $\delta_{\rm p\mu}=\omega_{\rm p\mu}^{0}-\omega_{21}$ and $\delta_{\rm c\mu}=\omega_{\rm c\mu}^{0}-\omega_{32}$.
Note that the DDI vanishes ($V^{\mu\nu}_{\rm e}=0$) when $\mu=\nu$. This means that Eqs.~(\ref{eq:alpha24AB})-(\ref{eq:alpha34AB}) correspond to the usual EIT Bloch equations \cite{fleischhauer_EIT_2005} in the individual cloud.


Finally, the 1D propagation equation for the probe field in clouds $A$ and $B$, obtained from the Maxwell equations \cite{scully_quantum_1997}, reads
%
\begin{eqnarray}
	\left(\partial_t+c\partial_z\right)\Omega_{\rm p\mu}\left(z,t\right) = ic\kappa_{\mu}\sum_{\nu=A,B}\int\alpha^{\mu\nu}_{24}\left(z,z',t\right)\alpha^{\nu}_{4}\left(z'\right)^{*}dz'. \label{eq:FieldProp}
\end{eqnarray}
The integrated expression at the right hand side depends on Eqs.~(\ref{eq:alpha24AB}) and (\ref{eq:alpha34AB}) and
corresponds to the macroscopic polarization, expressed in terms of the expectation value of the dipole operator at cloud $\mu$. The coupling constant $\kappa_{\mu}$ is defined as $\kappa_{\mu}=d_{12}^{2}\omega_{\rm p\mu}^{0}n_{\mu}/(2\epsilon_{0}\hbar c)$, with $c$ being the speed of light in vacuum, $n_{\mu}\equiv N_{\mu}/\mathcal{V}$ the atomic density, and $\mathcal{V}$ the quantization volume. The evolution equations Eqs.~(\ref{eq:alpha24AB})-(\ref{eq:FieldProp}) are linear in the probe field amplitudes $\Omega_{\rm p \mu}(z,t)$ and are in fact equivalent to those obtained with a quantum-field description, when interpreting the classical probe field as the single-photon probability amplitude \cite{Entangled_EIT_2014}. 

\subsection{Propagation equation for the PSS}
\label{sec:PropagationEquationForThePSS}

In order to solve the probe field propagation equation~(\ref{eq:FieldProp}), we first obtain $\alpha^{\mu\nu}_{24}$ through the adiabatic elimination~\cite{fleischhauer_EIT_2005}, i.e., we obtain the corresponding steady solution by forcing $\partial_{t}\alpha^{\mu\nu}_{24}=0$ in Eq.~(\ref{eq:alpha24AB}). 
Next, we solve Eqs.~(\ref{eq:alpha34AB})-({\ref{eq:FieldProp}}) in the frequency domain through the Fourier transform $\tilde{f}(\omega)\equiv\int{f(t)e^{-i\omega t}}dt$. This yields the propagation equation for the probe field amplitude
\begin{eqnarray}
	\partial_{z}\tilde{\Omega}_{\rm p\mu}\left(z,\omega\right) = i\chi^{\mu}_{L}\left(z,\omega\right)\tilde{\Omega}_{\rm p\mu}\left(z,\omega\right) +i\int\chi^{\mu}_{N}\left(z,z',\omega\right)\tilde{\Omega}_{\rm p\nu}\left(z',\omega\right)dz', \label{eq:fftFieldA}
\end{eqnarray}
where 
$\chi^\mu_L$ and $\chi^\mu_N$ are referred to as the local and non-local susceptibilities, respectively \cite{Entangled_EIT_2014}. 
We then define $\Omega_{{\rm c}\mu}(z)=\left|\Omega_{{\rm c}\mu}\right|e^{i(k_{\rm c}z+\varphi_{\rm c\mu})}$ and $\alpha^{\mu}_{4}(z)=\left|\alpha^{\mu}_{4}(z)\right|e^{ik_{\rm s}z}$, with $k_{\rm c}$ and $k_{\rm s}$ being the wavevectors of the control field and the initial spinwave excitation, Eq.~(\ref{eq:SpinwaveState}), respectively. 
After imposing equal parameters for the two clouds, i.e., $\delta_{{\rm p}A}=\delta_{{\rm p}B}=\delta_{\rm p}$, $\Delta_{{\rm p}A}=\Delta_{{\rm p}B}=\Delta_{\rm p}$, $\delta_{{\rm c}A}=\delta_{{\rm c}B}=\delta_{\rm c}$, $N_{A}=N_{B}=N$, $\rho_{A}=\rho_{B}=\rho$, $\kappa_{A}=\kappa_{B}=\kappa$, and $\left|\Omega_{{\rm c}A}\right|=\left|\Omega_{{\rm c}B}\right|=\Omega_{\rm c}$, the explicit forms of $\chi^\mu_L$ and $\chi^\mu_N$ read:
\begin{eqnarray}
	\chi^{\mu}_{L}\left(z,\omega\right)  = -\frac{\omega}{c}-\frac{\kappa}{\Delta_{\rm p}}-\kappa\frac{\Omega_{\rm c}^{2}}{\Delta_{\rm p}^2}\sum_{\nu=A,B}\int\frac{\left|\alpha^{\nu}_{4}\left(z'\right)\right|^{2}\Delta_{\rm s}(\omega)}{\Delta_s(\omega)^{2}-V^{\mu\nu}_{\rm e}\left(z,z'\right)^{2}}dz', \label{eq:Xp1L} \\
	\chi^{\mu}_{N}\left(z,z',\omega\right)  = -\kappa\frac{\Omega_{\rm c}^{2}}{\Delta_{\rm p}^2}\sum_{\nu=A,B}\frac{e^{i\phi_{\mu\nu}\left(z,z'\right)}\left|\alpha^{\nu}_{4}\left(z'\right)\alpha^{\mu}_{4}\left(z\right)\right|V^{\mu\nu}_{\rm e}\left(z,z'\right)}{\Delta_s(\omega)^{2}-V^{\mu\nu}_{\rm e}\left(z,z'\right)^{2}}. \label{eq:Xp1NL}
\end{eqnarray}
In the derivation of Eqs.~(\ref{eq:Xp1L})-(\ref{eq:Xp1NL}), we have abbreviated
\begin{eqnarray}
	\Delta_s(\omega)=\delta_{\rm p}+\delta_{\rm c}-\omega-\frac{\Omega_{\rm c}^{2}}{\Delta_{\rm p}}, \label{eq:ShiftedDetuning}
\end{eqnarray}
which determines the position and width of the absorption peaks in the EIT for the non-interacting case \cite{fleischhauer_EIT_2005}. 
Furthermore, $\phi_{\mu\nu}\left(z,z'\right)$ is the phase relation [in Eq.~(\ref{eq:Xp1NL})] between the phases of the control fields and that of the initial spinwave in each cloud, which reads
\begin{eqnarray}
	\phi_{AB}\left(z,z'\right)=-\phi_{BA}\left(z',z\right)=-(k_{\rm c}-k_{\rm s})(z-z')-\varphi_{AB}. \label{eq:PhaseMatching}
\end{eqnarray}
Here, $z$ and $z'$ refer to positions of atoms in clouds $\mu$ and $\nu$, respectively, and $\varphi_{AB}=\varphi_{{\rm c}A}-\varphi_{{\rm c}B}$. We have made the control field phase difference $\varphi_{AB}$ explicit in Eq.~(\ref{eq:PhaseMatching}), since it allows to control the PSS filtering in our setup as we will show in the next section. 

The propagation equation for the PSS single photon, Eq.~(\ref{eq:fftFieldA}), is one of the central results of this work. It shows that the single-photon propagation is determined by the usual local susceptibility $\chi^\mu_L$ but also by a non-local part $\chi^\mu_N$ \cite{Entangled_EIT_2014}. The latter leads, by virtue of the DDI, to a coupling of the photon dynamics between the two clouds. In the following we will demonstrate that the non-local susceptibility renders exotic photon dynamics.

\subsection{Approximate analytical solution}
\label{sec:ApproximateAnalyticalSolution}

In order to obtain a qualitative understanding of the light propagation, we derive an analytical solution of Eq.~(\ref{eq:fftFieldA}) by carrying out additional approximations. First, we approximate the DDI profile by a square function $\widetilde{V}^{AB}_{\mathrm{e}}$ [see Fig.~\ref{fig:fig1}(c)] of height $V_{0}$ and width equal to the FWHM of the actual DDI strength $2z_{d}$. 
Next, we perform a local-field approximation \cite{Entangled_EIT_2014,sevincli_nonlocal_2011}, i.e., we take the field $\tilde{\Omega}_{\rm p\nu}\left(z',\omega\right)$ out of the integral in Eq.~({\ref{eq:fftFieldA}}) by assuming that the shape of the probe pulse does not change significantly within $2z_{d}$ and neglecting boundary effects. 
Moreover, for simplicity we impose the condition $k_{\rm c}-k_{\rm s}=0$, which can be obtained by properly aligning the spinwave excitation lasers and the control fields. This leads to $\phi_{\mu\nu}\left(z,z'\right)=-\varphi_{\mu\nu}$, according to Eq.~(\ref{eq:PhaseMatching}). Finally, we Taylor expand Eqs.~(\ref{eq:Xp1L})-(\ref{eq:Xp1NL}) up to first order in $\omega$, i.e., we assume a narrow-band probe pulse.
With these approximations, and defining the boundary conditions for the field as $\tilde{\Omega}_{\rm p\mu}(z=0,\omega)=\tilde{\Omega}^{(0)}_{\rm p\mu}(\omega)$, the solutions of the propagation equation~({\ref{eq:fftFieldA}}) are
\begin{eqnarray}
	\tilde{\Omega}_{{\rm p}A}\left(z,\omega\right)  = & \frac{\tilde{\Omega}^{(0)}_{{\rm p}A}\left(\omega\right)+e^{-i\varphi_{AB}}\tilde{\Omega}^{(0)}_{{\rm p}B}\left(\omega\right)}{2}e^{i\left(\eta_{+}-\frac{\omega}{v_{+}}\right)z} \nonumber \\
	& + \frac{\tilde{\Omega}^{(0)}_{{\rm p}A}\left(\omega\right)-e^{-i\varphi_{AB}}\tilde{\Omega}^{(0)}_{{\rm p}B}\left(\omega\right)}{2}e^{i\left(\eta_{-}-\frac{\omega}{v_{-}}\right)z}, \label{eq:OmegaAsol} \\
	\tilde{\Omega}_{{\rm p}B}\left(z,\omega\right)  = & \frac{e^{i\varphi_{AB}}\tilde{\Omega}^{(0)}_{{\rm p}A}\left(\omega\right)+\tilde{\Omega}^{(0)}_{{\rm p}B}\left(\omega\right)}{2}e^{i\left(\eta_{+}-\frac{\omega}{v_{+}}\right)z} \nonumber \\
	& -\frac{e^{i\varphi_{AB}}\tilde{\Omega}^{(0)}_{{\rm p}A}\left(\omega\right)-\tilde{\Omega}^{(0)}_{{\rm p}B}\left(\omega\right)}{2}e^{i\left(\eta_{-}-\frac{\omega}{v_{-}}\right)z}, \label{eq:OmegaBsol}
\end{eqnarray}
where
\begin{eqnarray}
\eta_{\pm} = & -\frac{\kappa}{\Delta_{\rm p}}-\kappa\frac{\Omega_{\rm c}^{2}}{\Delta_{\rm p}^2}\left[\frac{z_{d}\rho}{\Delta_{\rm s}(0)\mp V_{0}}+\frac{1-z_{d}\rho}{\Delta_{\rm s}(0)}\right], \label{eq:etapm} \\
v_{\pm}^{-1} = & \frac{1}{c}+\kappa\frac{\Omega_{\rm c}^{2}}{\Delta_{\rm p}^2}\left[\frac{z_{d}\rho}{\left[\Delta_{\rm s}(0)\mp V_{0}\right]^{2}}+\frac{1-z_{d}\rho}{\Delta_{\rm s}(0)^{2}}\right]. \label{eq:velocpm}
\end{eqnarray}
For simplicity the single-particle linear density $\rho(z)$ has been considered constant along the medium, i.e., $\rho=1/L$.
The two terms inside the brackets at the right hand side of Eqs.~(\ref{eq:etapm})-(\ref{eq:velocpm}) are important to determine the light propagation. Here the first term depends on the interaction strength $V_0$ between the atoms belonging to different clouds, while the second one corresponds to the conventional EIT contribution found in interaction-free atomic gases. 

Assuming $\delta_{\rm p}\gg\gamma$, we obtain the analytical solution of the photon field by performing the inverse Fourier transform of Eqs.~(\ref{eq:OmegaAsol})-(\ref{eq:OmegaBsol})
\begin{eqnarray}
	\Omega_{\pm}\left(z,t\right) = e^{i\eta_{\pm}z}\Omega^{(0)}_{\pm}\left(t-\frac{z}{v_{\pm}}\right). \label{eq:SymAntisymField}
\end{eqnarray}
Here $\Omega_{+}$ and $\Omega_{-}$ are the symmetric and antisymmetric superpositions (or normal modes, by analogy with the results in \cite{Harris_NormalModes_1994}) of the probe field in the two clouds, defined as
\begin{eqnarray}
	\Omega_{\pm}\left(z,t\right)\equiv\Omega_{{\rm p}A}\left(z,t\right)\pm e^{-i\varphi_{AB}}\Omega_{{\rm p}B}\left(z,t\right). \label{eq:SymAntisymSuperpos}
\end{eqnarray}
These normal modes form an orthogonal basis, whose specific form depends on the phase difference between the control fields $\varphi_{AB}$, allows us to describe the photon propagation in a simple manner. In particular, from Eq.~(\ref{eq:SymAntisymField}) we see that the quantities ${\rm Im}[\eta_{\pm}]$ and $v_{\pm}$ defined in Eqs.~(\ref{eq:etapm})-(\ref{eq:velocpm}) are, respectively, the absorption coefficient and propagation velocity of the normal modes, Eq.~(\ref{eq:SymAntisymSuperpos}), inside the medium. 
%

\begin{figure}
\centering
\includegraphics*[width=1\columnwidth]{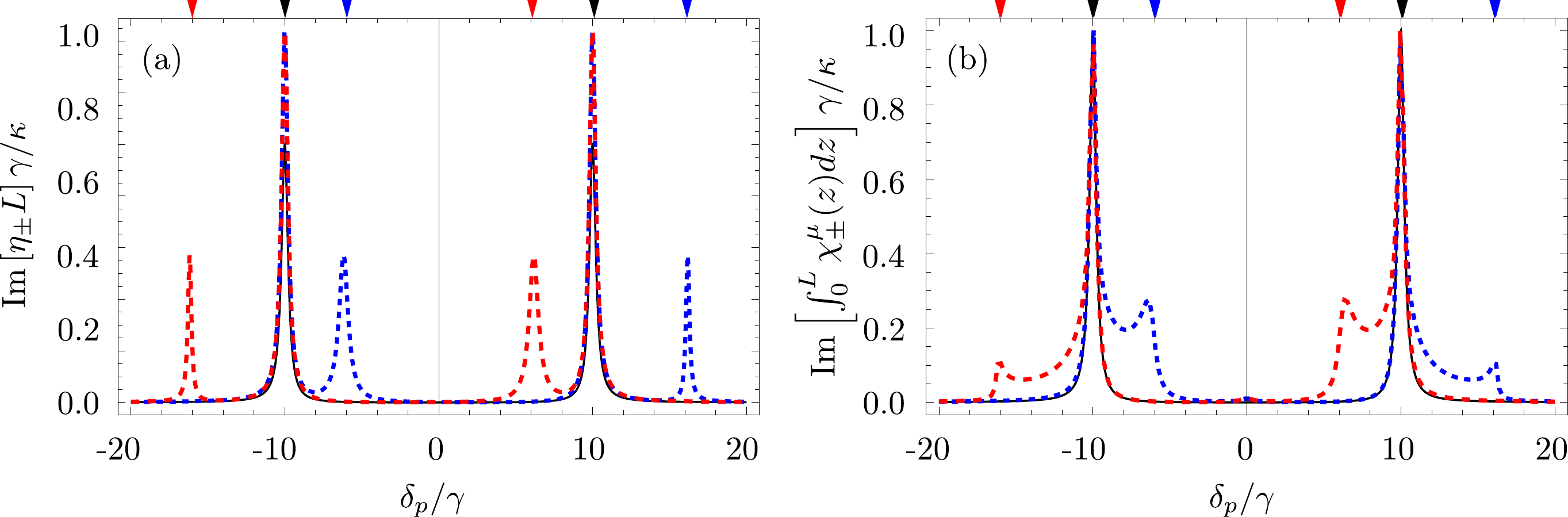}
\caption{Imaginary part of the optical depth (a) $\eta_{\pm}L$ and (b) $\int^L_0{X^{\mu}_{\pm}(z)}dz$, rescaled by $\gamma/\kappa$, as a function of the probe detuning $\delta_{\rm p}/\gamma$. Blue and red dashed lines correspond to optical response of the atoms to the symmetric and antisymmetric probe field component, respectively. The non-interacting case ($V_{0}=0$) is shown with a black line, and rescaled to half of the original height for better comparison.
The parameters are $\delta_{\rm c}=0$, $V_{0}=10\gamma$, $\ell=0.5L$, and $\Omega_c=10\gamma$.}
\label{fig:fig2}
\end{figure}

Therefore, the solutions presented in Eq.~(\ref{eq:SymAntisymField}) describe the propagation of two orthogonal PSS for the single photon whose absorption/transmission and refractive properties can be controlled by tuning the system parameters.
To exemplify this, we plot in Fig.~\ref{fig:fig2}(a) the imaginary part of $\eta_{\pm}L$ (i.e., the optical depth for the $\Omega_{\pm}$ solutions), with blue and red dashed lines, respectively, as a function of the probe detuning $\delta_{\rm p}$, for $\delta_{\rm c}=0$.
When $V_{0}=0$ one recovers the usual transparency window of the EIT \cite{fleischhauer_EIT_2005} (shown with black solid line) since $\eta_{+}=\eta_{-}$, and thus both components $\Omega_{\pm}$ are equally absorbed.
In the presence of DDI this feature of non-interacting EIT is still visible (see blue and red peaks marked by black arrows). 
However, the interaction term in [see Eq.~(\ref{eq:etapm})] gives rise to two additional pairs of EIT peaks (blue and red dashed peaks marked by correspondingly colored arrows). Those are shifted in opposite directions according to the sign of the interaction strength $\pm V_0$, which acts as an effective detuning on top of the control field detuning. 

For comparison, we also consider an approximated solution for Eq.~(\ref{eq:fftFieldA}) using the actual DDI potential, Eq.~(\ref{eq:Ve}), and again the local-field approximation. For this situation we can still express the solutions for the field propagation in terms of the symmetric and antisymmetric modes
\begin{eqnarray}
	\tilde{\Omega}_{\pm}\left(L,\omega\right) = \exp\left[i\int^{L}_{0}{X^{\mu}_{\pm}(z)}dz\right]\tilde{\Omega}^{(0)}_{\pm}\left(\omega\right), \label{eq:SymAntisymFieldReal} \\
	X^{\mu}_{\pm}(z)=\chi^{\mu}_{L}\left(z,0\right)\pm\int^{L}_{0}{\chi^{\mu}_{N}\left(z,z',0\right)}dz'.  \label{eq:IntSuscep}
\end{eqnarray}
Since we are only interested in the absorption of the medium we have approximated $X^{\mu}_{\pm}(z)$ to zeroth order in $\omega$. By doing so, we assume that the medium response is the same for all the photon frequency components, i.e, we consider the continuous wave regime, and we neglect dispersion effects. The integral in the exponent at the right hand side of Eq.~(\ref{eq:SymAntisymFieldReal}) determines the total absorption of the medium for the probe light, i.e., the optical depth. 
For comparison with the data shown in Fig.~\ref{fig:fig2}(a), we plot in Fig~\ref{fig:fig2}(b) the imaginary part of this quantity $\int^{L}_{0}{X^{\mu}_{\pm}(z)}dz$, with blue and red dashed lines for the $+$ and $-$ cases, respectively. The integral is performed numerically, using the same parameters as in Fig~\ref{fig:fig2}(a). We observe the same peak positions as in Fig.~\ref{fig:fig2}(a), indicated by colored arrows (the non-interacting case is again presented with a black solid line). However, we notice that those peaks which are shifted from the non-interacting resonance (red and blue arrows) are widened and their height has decreased with respect to the peaks obtained from the square potential approach. This difference is caused by the variation of the interaction potential depending on the different interatomic distances, which effectively acts as a position dependent control detuning. Thus, the integration of the susceptibility over $z$ produces an inhomogeneous broadening around the interaction peaks (blue and red arrows). This broadening is not seen around the peaks indicated by black arrows, because for such probe detuning $\delta_{p}$ the contribution from atoms being in the same cloud dominates over the contribution from interacting atoms.

From Fig.~\ref{fig:fig2}, we realize that the opposite shifts experienced by the red and blue peaks give rise to a large difference between the absorption coefficients of the two normal modes in Eq.~(\ref{eq:SymAntisymSuperpos}). Thus the probe detuning $\delta_{\rm p}$ represents a handle to control the transmission and absorption of PSS. For instance, if we tune our system with a probe detuning $\delta_{\rm p}$ close to any of the blue arrows, the symmetric mode $\Omega_{+}$ will experience a much higher absorption than the antisymmetric superposition $\Omega_{-}$, rendering the former the ``bright state'' and the latter the ``dark state''. 
An example of this situation will be demonstrated numerically in the next section. 
If on the contrary the system is tuned close to one of the red shifted peaks, the role of ``bright'' and ``dark'' state will be swapped. Moreover, we have also the additional control parameter, the phase difference of the control fields $\varphi_{AB}$, which rotates the basis that defines the symmetric and antisymmetric normal modes in Eq.~(\ref{eq:SymAntisymSuperpos}). This parameter then allows to choose which is the specific form of the superposition state at the output of the medium. 


\subsection{Numerical results}
\label{sec:NumericalResults}

In the following section we perform numerical simulations of the light propagating in the two clouds considering the actual interaction potential, given by expression Eq.~(\ref{eq:Ve}), and compare the results with the analytical solution.
For the simulations, we use a spatio-temporal grid to propagate the optical-Bloch equations, Eqs.~(\ref{eq:alpha24AB})-(\ref{eq:FieldProp}) and assume the probe Gaussian pulse much wider than the length of the medium $L$. For the atoms, we have used parameters corresponding to states $|3\rangle\equiv|90S,J=m_J=1/2\rangle $ and $|4\rangle\equiv|90P,J=m_J=1/2\rangle$ of $^{87}$Rb. 
The remaining parameters are: $V_0=10\gamma$, $\ell=0.5L$, $\kappa=9\gamma/L$, $\Omega_c=10\gamma$, $\delta_p=-6.5\gamma$, and $\delta_c=0$, with $\gamma\simeq38$ MHz and $L=15\,\mu$m, similar to those used in magneto-optical trap experiments \cite{peyronel_2012,dudin_observation_2012,maxwell_storage_2013}.
These parameters correspond to the blue peaks in Figs.~\ref{fig:fig2}(a)-(b), marked by the blue arrows at negative probe detuning $\delta_{\rm p}$. 

\begin{figure}
\centering
\includegraphics*[width=0.6\columnwidth]{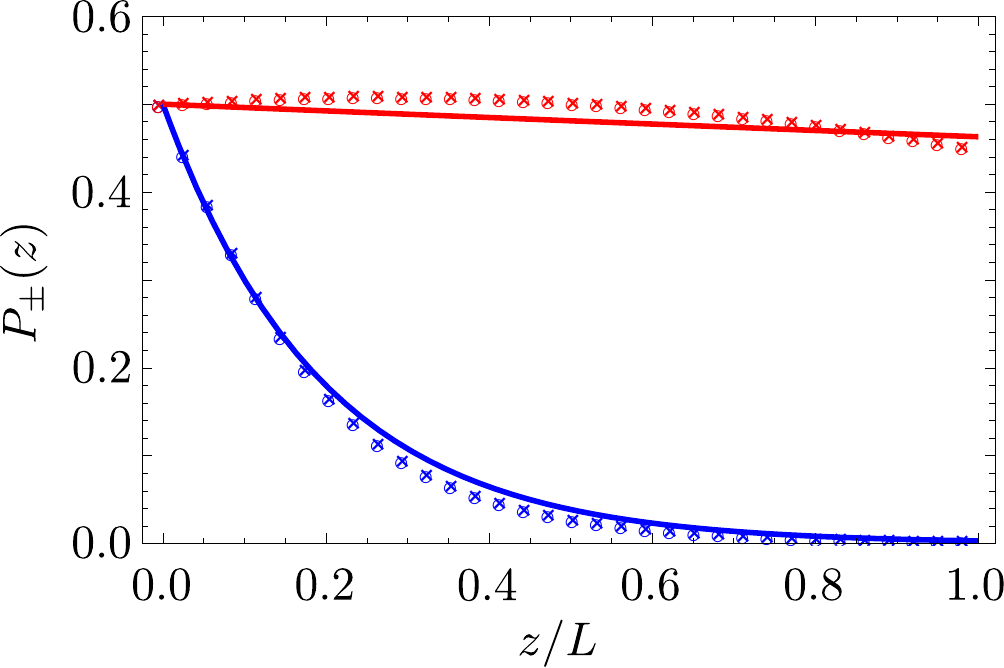}
\caption{Spatial profile of the probability for the photon being in the symmetric (blue) or antisymmetric (red) superposition state, for $\varphi_{AB}=0$, $\Omega^{(0)}_{{\rm p}B}(t)=0$ (circles), and $\varphi_{AB}=\pi/2$, $\Omega^{(0)}_{{\rm p}B}(t)=\Omega^{(0)}_{{\rm p}A}(t)$ (crosses). The remaining parameters are $V_0=10\gamma$, $\ell=0.5L$, $\kappa=9\gamma/L$, $\Omega_c=10\gamma$, $\delta_p=-6.5\gamma$, and $\delta_c=0$. The solid lines correspond to the analytical solution using the values from Fig.~\ref{fig:fig2}(b).}
\label{fig:fig3}
\end{figure}

For the simulations we select two situations corresponding to different initial PSS entering the medium. First, we consider the case where the photon is only entering cloud $A$, i.e., $\int{|\Omega^{(0)}_{{\rm p}A}(t)|^2|}dt=1$ and $\int{|\Omega^{(0)}_{{\rm p}B}(t)|^2|}dt=0$, and the relative phase between the control fields is $\varphi_{AB}=0$. This means that the input state in the normal mode basis is $\Omega^{(0)}_{\pm}(t)=\Omega^{(0)}_{{\rm p}A}(t)$.
In Fig.~\ref{fig:fig3}  we plot the probability $P_{\pm}(z)$ of finding the photon in the symmetric (blue circles) or antisymmetric (red circles) superposition states, defined as
\begin{eqnarray}
		P_{\pm}(z)=\frac{\int{\left|\Omega_{\pm}\left(z,t\right)\right|^{2}dt}}{\int{\left[\left|\Omega^{(0)}_{+}\left(t\right)\right|^2+\left|\Omega^{(0)}_{-}\left(t\right)\right|^2\right]dt}}, \label{eq:ProbPM}
\end{eqnarray}
where $\Omega^{(0)}_{\pm}\left(t\right)$ are defined in Eqs.~(\ref{eq:SymAntisymField})-(\ref{eq:SymAntisymSuperpos}). The analytical result is also plotted with solid lines, using Eqs.~(\ref{eq:SymAntisymFieldReal})-(\ref{eq:IntSuscep}). 
From the figure we observe that during propagation the symmetric superposition $\Omega_{+}$ is strongly absorbed compared to the  antisymmetric component $\Omega_{-}$. Thus the behavior of the light intensity in the medium is well described by the analytical model. 
The second situation considered corresponds to the propagation of two identical path-components with no initial phase difference between them, i.e., $\Omega^{(0)}_{{\rm p}B}(t)=\Omega^{(0)}_{{\rm p}A}(t)$ such that $\int{|\Omega^{(0)}_{{\rm p}A}(t)|^2|}dt=\int{|\Omega^{(0)}_{{\rm p}B}(t)|^2|}dt=1/2$ and $\int{\arg[\Omega^{(0)}_{{\rm p}B}(t)\Omega^{(0)}_{{\rm p}A}(t)^{*}]}dt=0$. 
For the phase difference between the control fields we choose $\varphi_{AB}=\pi/2$. The corresponding probabilities are shown in the same figure~\ref{fig:fig3} with blue and red crosses. For this situation the results are again in agreement with the analytical model. 
In both situations presented in Fig.~\ref{fig:fig3}, the small discrepancies found in the absorption profile 
are mainly due to boundary effects, which have been neglected in Eq.~(\ref{eq:SymAntisymFieldReal}). 
Thus this system can be used to filter a particular PSS.

Finally, we would like to note that similar light propagation effects have been reported in the context of coherently coupled atomic systems. In particular, the propagation of matched pulses \cite{Harris_NormalModes_1994,Konopnicki_Simultons_1981,Harris_MatchedEIT_1993,Eberly_DressedField_1994,Grobe_Adiabatons_1994,Fleischhauer_Pulsematching_1995,Xiao-xue_PropagatingEntangledState_2005,Viscor_Phaseonium_2011} has been widely investigated.
In this propagation effect, two different polarization \cite{Eberly_DressedField_1994,Viscor_Phaseonium_2011} or frequency \cite{Konopnicki_Simultons_1981,Fleischhauer_Pulsematching_1995,Xiao-xue_PropagatingEntangledState_2005} components which are coupled to adjacent atomic transitions are able to interchange energy as they propagate. 
In general, different conditions have to be satisfied in order to observe this phenomenon. First, the two light components must be coupled to a common bare state of, for instance, $\Lambda$, $V$, or double-$\Lambda$ systems, which provides a path for the light to be exchanged. Also, an atomic coherence between the bare states that involve the two-photon transition is required. This atomic coherence is established by the propagating pulses themselves when they are intense enough, investing part of their energy in creating the dark atomic superposition state. However, for weak or single-photon pulses this coherence must be created beforehand, e.g., by preparing the medium in a phaseonium state \cite{Scully_phaseonium_1992,Viscor_Phaseonium_2011}. In all the cases, once this coherence is built up, the light components evolve into a dark superposition state, sometimes called simulton \cite{Konopnicki_Simultons_1981} or adiabaton \cite{Grobe_Adiabatons_1994}, that propagates transparently, while the orthogonal superposition is absorbed into the medium. 
Analogously, we have considered here a system prepared initially in a certain coherent state, i.e., the initial spinwave Eq.~(\ref{eq:SpinwaveState}), and a two-component light pulse, in a PSS, propagating through the medium. 
In our system, the exchange interaction creates pair states of the form $|\pm\rangle=|3\rangle_{i_{A}}|4\rangle_{j_{B}}\pm|4\rangle_{i_{A}}|3\rangle_{j_{B}}$, so called excitonic states \cite{Entangled_EIT_2014}. These are the common states in our situation that allows the light to be transferred from one cloud to another. This is the reason for which we observe eventually matched pulse propagation. However, the novelty here with respect to previous studies is that, for the first time to our knowledge, we describe matched propagation of components that are delocalized in space.

\section{Generalization to $\mathcal{N}$ clouds}
\label{sec:GeneralizationToNClouds}

In this section we discuss a generalization of the previous setup from two to $\mathcal{N}$ parallel and equally spaced clouds. 
We consider DDI to be present only between neighboring clouds. In this situation, the equations describing the evolution of the system are the formally equivalent to Eqs.~(\ref{eq:alpha24AB})-(\ref{eq:alpha34AB}) with the difference that now the subindexes $\mu,\nu=\{1,2,3,...,\mathcal{N}\}$ run over all the possible clouds. In addition, the interaction strength is now $\tilde{V}^{\mu\nu}_{\rm e}=V^{\mu\nu}_{\rm e}(\delta_{\mu\nu+1}+\delta_{\mu\nu-1})$.
With this premise, and making use of the same approximations that led to Eqs.~(\ref{eq:OmegaAsol})-(\ref{eq:OmegaBsol}), the system of equations for the field components propagating in $\mathcal{N}$-cloud system, Eq.~(\ref{eq:fftFieldA}), can be written as
\begin{eqnarray}
	\partial_{z}\Omega_{\rm p\mu} = i\chi_D\Omega_{\rm p\mu}+i\chi_S\left(\Omega_{\rm p\mu-1}+\Omega_{\rm p\mu+1}\right). \label{eq:PropEqMulti}
\end{eqnarray}
Here
\begin{eqnarray}
	\chi_D = -\frac{\kappa}{\Delta_{\rm p}}-\kappa\frac{\Omega_{\rm c}^{2}}{\Delta_{\rm p}^2}\left[\frac{\Delta_{\rm s}(0)4z_{d}\rho_{\mathcal{N}}}{\Delta_{\rm s}(0)^2- V_{0}^2}+\frac{1-4z_{d}\rho_{\mathcal{N}}}{\Delta_{\rm s}(0)}\right], \label{eq:DiagCoeff} \\
	\chi_S = -\kappa\frac{\Omega_{\rm c}^{2}}{\Delta_{\rm p}^2}\frac{V_{0}2z_{d}\rho_{\mathcal{N}}}{\Delta_{\rm s}(0)^{2}-V_{0}^{2}}, \label{eq:SubSupDiagCoeff}
\end{eqnarray}
are a generalization of the local and non-local susceptibilities in Eqs.~(\ref{eq:Xp1L})-(\ref{eq:Xp1NL}), respectively, with $\rho_{\mathcal{N}}\equiv\rho/\mathcal{N}$ being the single particle density per cloud. 

In order to gain a first understanding of the behavior that one can expect for the light propagation in the $\mathcal{N}$-cloud system let us note the analogy between Eq.~(\ref{eq:PropEqMulti}) and the 1D Schr\"odinger equation for a free particle. 
This becomes manifest when considering $\mathcal{N}\gg1$, and a transverse smooth variation of the probe photon field over the intercloud distance $\ell$. Under these considerations 
and using the discretization of the Laplacian, $\partial_{y}^2f(y)\simeq\left[f(y+\ell)-2f(y)+f(y-\ell)\right]/\ell^2$ with $y$ the continuous version of the cloud position $\mu\ell$, Eq.~(\ref{eq:PropEqMulti}) can  be rewritten as
\begin{eqnarray}
	i\hbar\partial_{z}\Omega_{\rm p}(y,z)=-\left[\frac{\hbar^2}{2m}\partial^{2}_{y}+\Gamma\right]\Omega_{\rm p}(y,z).
\label{eq:Difusion}
\end{eqnarray}
%
Here $\Gamma=\hbar\left(\chi_D+2\chi_S\right)$ is a complex number whose imaginary part $\Gamma_{\rm i}$ leads to a reduction of the norm of the probe field, i.e, absorption. 
Note that the propagation distance $z$ takes the role of time and the parameter $m^{-1}=m^{-1}_{\rm r}+im^{-1}_{\rm i}=2\chi_S\ell^2/\hbar$ can be formally regarded as a complex mass. This analogy already anticipates that a photon entering one of the clouds will spread or diffuse among neighboring clouds, in the transverse direction $y$, and will decay due to the imaginary parts of the mass $m_{\rm i}$ and $\Gamma_{\rm i}$. 
We analytically demonstrate these effects by assuming an initial transverse profile of the probe field to be a Gaussian distribution of the form: $\Omega_{\rm p}(y,0)=\exp\{-y^2/[2\sigma(0)]^2\}/[2\pi\sigma(0)^2]^{1/4}$, with $\int{|\Omega_{\rm p}(y,0)|^2dy}=1$.
By solving Eq.~(\ref{eq:Difusion}), we obtain the variation of the width of the Gaussian transverse profile $\sigma(z)$ as well as its norm $h(z)$ as a function of the propagation distance $z$:
\begin{eqnarray}
	h(z) = \frac{\exp{\left[2\Gamma_{\rm i}z\right]}}{\sqrt{1+\frac{z}{2m_{\rm i}\sigma(0)^2}}}, \label{eq:Norm} \\
	\sigma(z)^2\equiv\frac{\left\langle y^2\right\rangle}{h(z)}=\sigma(0)^2+\frac{z}{2m_{\rm i}}\left\{1+\frac{\frac{z}{m_{\rm i}}}{\frac{m_{\rm r}^2}{m_{\rm i}^2}\left[\frac{z}{m_{\rm i}}+2\sigma(0)^2\right]}\right\}. \label{eq:Width}
\end{eqnarray}
These results indicate that the norm $h(z)$ of the Gaussian and thus the probe photon intensity decrease with the propagation distance. Moreover we find that the probe photon will diffuse across the cloud since $\sigma(z)$ increases. Note that for $m_{\rm i}^{-1}\rightarrow0$, i.e., when $\delta_{\rm p}\gg\gamma$, one recovers the dynamics of a free particle in an initially Gaussian wavepacket state.

\begin{figure}
\centering
\includegraphics*[width=1\columnwidth]{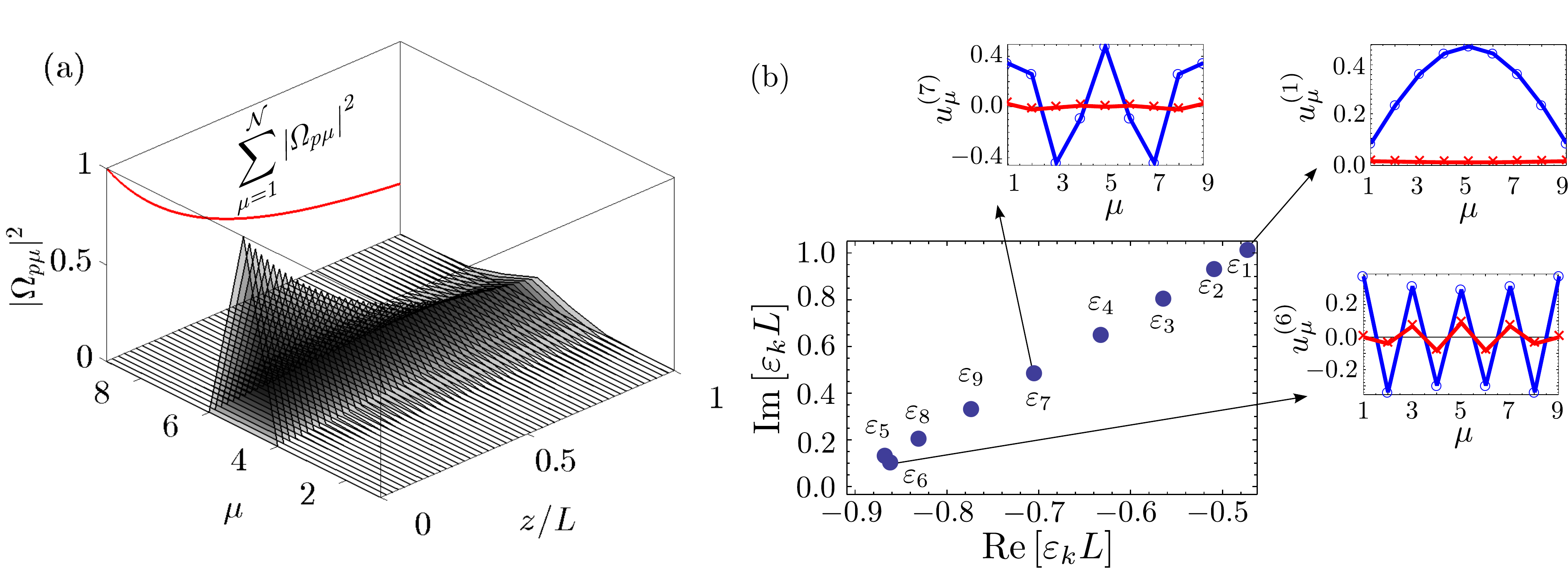}
\caption{(a) Light intensity in $\mathcal{N}=9$ clouds as a function of the propagation distance $z/L$. The photon is entering cloud 5 at $z=0$ and decays during propagation while diffusing across neighboring clouds. In the back panel we show the sum of the probe field intensities of all the clouds. (b) Complex plane with the different eigenvalues of the operator $\mathcal{X}$ [see Eq.~(\ref{eq:PropEqMultiMatrix})]. For selected eigenvalues, indicated by an arrow, we plot the real (blue circles) and imaginary (red crosses) part of the corresponding eigenvector components.}
\label{fig:fig4}
\end{figure}

On the other hand, for a quantitative description of the system one needs to solve the discrete system of coupled equations (\ref{eq:PropEqMulti}) numerically. In a matrix form they read, 
\begin{eqnarray}
	\partial_{z}{\bf W}=i\mathcal{X}{\bf W}, \label{eq:PropEqMultiMatrix}
\end{eqnarray}
where ${\bf W}=(\Omega_{\rm p1},...,\Omega_{\rm p\mu},...,\Omega_{\rm p\mathcal{N}})^{t}$ and the coefficient matrix $\mathcal{X}$ is a symmetric tridiagonal matrix with elements $\chi_\mathcal{D}$ in the diagonal and elements $\chi_\mathcal{S}$ in the upper and lower diagonals. Then the eigenvalues $\varepsilon_k$ and eigenvectors ${\bf u}^{k}$ of $\mathcal{X}$ determine the solution of the propagation equation.
In the following we briefly discuss a numerical example with $\mathcal{N}=9$ clouds and the single photon entering initially the middle cloud $\mu=5$. In Fig.~\ref{fig:fig4}(a), we show the solution for the probe light intensity propagating in the different clouds $\mu$ as a function of the propagation distance, using the same parameters as in Fig.~\ref{fig:fig2}. Clearly, one observes light diffusion as predicted from Eq.~(\ref{eq:Difusion}). 
Moreover, the decay of the intensity during propagation is consistent with the fact that all the eigenvalues $\varepsilon_{k}$ 
have positive imaginary part. This is shown in Fig.~\ref{fig:fig4}(b), where the different $\varepsilon_{k}$'s are represented in the complex plane. Furthermore,
The real (blue circles) and imaginary (red crosses) parts of some selected eigenvectors are shown in the insets of Fig.~\ref{fig:fig4}(b) as a function of the component index $\mu$. 
The modes whose eigenvalues have a large imaginary part, e.g., upper-right inset of Fig.~\ref{fig:fig4}(b) with $\varepsilon_{1}$, exhibit a stronger decay than those with ${\rm Im}[\varepsilon_{k}]\simeq0$, e.g., lower-right inset of Fig.~\ref{fig:fig4}(b) with $\varepsilon_{6}$. Focusing on the total light intensity, i.e, $\sum_{\mu}\left|\Omega_{\rm p\mu}\right|^2$ [see back panel in Fig.~\ref{fig:fig4}(a)], the contribution of the latter eigenvalues becomes more apparent at large propagation distances, since the fast decaying modes are absorbed  already at short propagation distances. Close to the end of the medium $(z\simeq L)$ the overall absorption rate of the propagating light is small, leading to a relatively slow decay of the probe intensity. 

\section{Summary and conlusions}
\label{sec:SummaryAndConlusions}

In this work we have studied the propagation of a single-photon wavepacket in a path-superposition state (PSS) through dipole-dipole coupled atomic clouds under conditions of EIT.
We have considered that before the photon enters the clouds they are prepared in a collectively excited Rydberg state $nP_{1/2}$. 
Then, when the single photon propagates in an EIT configuration it excites a second Rydberg state $nS_{1/2}$, which is coupled with the first one via the DDI. This coherent coupling provides population exchange between the Rydberg states, which causes the photon being transferred between different clouds.

We have focused on the situation where the angle of the dipoles with respect to the propagation axis allows excitation exchange only between atoms of the different clouds. In case of two parallel atomic clouds, the photon propagates as a combination of symmetric and antisymmetric PSS with different absorptive and dispersive properties. In a certain parameter regime we have shown that the antisymmetric PSS can be regarded as a quasi-dark state, experiencing significantly weaker absorption compared to the symmetric PSS (bright state).
Under these conditions, the system effectively behaves as a PSS filter that can be tuned using the phase between the control fields of the different clouds.
We have provided analytical solutions for the probe light propagation in order to shed light on the physical mechanisms behind the formation of the bright and dark spatially propagating modes. The analytical model has been verified by numerical simulations of the light propagation and the ability of the system to filter PSS has been confirmed.
Finally, we have studied the generalized case of $\mathcal{N}$ nearest-neighbor coupled clouds. We have obtained an analytical model in which, analogously to the one-dimensional Schr\"odinger equation for a free particle with a complex mass, the light entering the medium spreads among the clouds and decays during propagation. This behavior has been confirmed by numerically studying a particular case with $\mathcal{N}=9$ clouds.

Nevertheless, the full potential of our system has not been thoroughly exploited in this work. We anticipate that varying the control fields (their strength and detuning) that couple each cloud would provide extra tunability over the final PSS. For instance, by adiabatically turning off and on the control fields \cite{maxwell_storage_2013} during the photon propagation, one could store the dark PSS in the (long lived) ground-Rydberg coherence and recover it at a later time \cite{Choi_Mapping_2008}. In conjunction with these storage and on-demand releasing techniques of photon states~\cite{fleischhauer_EIT_2005}, our study paves an alternative approach towards quantum communication and information processing applications with strong and nonlocal Rydberg interactions.


\section{Acknowledgments}
\label{sec:Acknowledgments}

The research leading to these results has received funding from the European Research Council under the European Union's Seventh Framework Programme (FP/2007-2013) / ERC Grant Agreement No. 335266 (ESCQUMA), the EU-FET grants HAIRS 612862 and QuILMI 295293 as well as from the University of Nottingham.

\section{References}

\bibliography{ReferenceLibrary}

\providecommand{\newblock}{}
\begin{thebibliography}{10}
\expandafter\ifx\csname url\endcsname\relax
  \def\url#1{{\tt #1}}\fi
\expandafter\ifx\csname urlprefix\endcsname\relax\def\urlprefix{URL }\fi
\providecommand{\eprint}[2][]{\url{#2}}

\bibitem{saffman_quantum_2010}
Saffman M, Walker T~G and M{\o}lmer K 2010 {\em Rev. Mod. Phys.\/} {\bf 82}
  2313 \urlprefix\url{http://link.aps.org/doi/10.1103/RevModPhys.82.2313}

\bibitem{fleischhauer_EIT_2005}
Fleischhauer M, Imamoglu A and Marangos J~P 2005 {\em Rev. Mod. Phys.\/} {\bf
  77}(2) 633--673
  \urlprefix\url{http://link.aps.org/doi/10.1103/RevModPhys.77.633}

\bibitem{gorshkov_dissipative_2013}
Gorshkov A~V, Nath R and Pohl T 2013 {\em Phys. Rev. Lett.\/} {\bf 110} 153601
  \urlprefix\url{http://link.aps.org/doi/10.1103/PhysRevLett.110.153601}

\bibitem{pritchard_cooperative_2010}
Pritchard J~D, Maxwell D, Gauguet A, Weatherill K~J, Jones M~P~A and Adams C~S
  2010 {\em Phys. Rev. Lett.\/} {\bf 105} 193603
  \urlprefix\url{http://link.aps.org/doi/10.1103/PhysRevLett.105.193603}

\bibitem{peyronel_2012}
Peyronel T, Firstenberg O, Liang Q~Y, Hofferberth S, Gorshkov A~V, Pohl T,
  Lukin M~D and Vuleti\'{c} V 2012 {\em Nature\/} {\bf 488} 57--60
  \urlprefix\url{http://dx.doi.org/10.1038/nature11361}

\bibitem{Tiarks_transistor_2014}
Tiarks D, Baur S, Schneider K, D\"urr S and Rempe G 2014 {\em Phys. Rev.
  Lett.\/} {\bf 113}(5) 053602
  \urlprefix\url{http://link.aps.org/doi/10.1103/PhysRevLett.113.053602}

\bibitem{Gorniaczyk_transistor_2014}
Gorniaczyk H, Tresp C, Schmidt J, Fedder H and Hofferberth S 2014 {\em Phys.
  Rev. Lett.\/} {\bf 113}(5) 053601
  \urlprefix\url{http://link.aps.org/doi/10.1103/PhysRevLett.113.053601}

\bibitem{baur_single_2014}
Baur S, Tiarks D, Rempe G and D\"urr S 2014 {\em Phys. Rev. Lett.\/} {\bf
  112}(7) 073901
  \urlprefix\url{http://link.aps.org/doi/10.1103/PhysRevLett.112.073901}

\bibitem{he_two-photon_2014}
He B, Sharypov A, Sheng J, Simon C and Xiao M 2014 {\em Phys. Rev. Lett.\/}
  {\bf 112} 133606
  \urlprefix\url{http://link.aps.org/doi/10.1103/PhysRevLett.112.133606}

\bibitem{Bienias_Scattering_2014}
Bienias P, Choi S, Firstenberg O, Maghrebi M~F, Gullans M, Lukin M~D, Gorshkov
  A~V and B\"uchler H~P 2014 {\em Phys. Rev. A\/} {\bf 90}(5) 053804
  \urlprefix\url{http://link.aps.org/doi/10.1103/PhysRevA.90.053804}

\bibitem{Honer_substractor_2011}
Honer J, L\"ow R, Weimer H, Pfau T and B\"uchler H~P 2011 {\em Phys. Rev.
  Lett.\/} {\bf 107}(9) 093601
  \urlprefix\url{http://link.aps.org/doi/10.1103/PhysRevLett.107.093601}

\bibitem{gorshkov_photon-photon_2011}
Gorshkov A~V, Otterbach J, Fleischhauer M, Pohl T and Lukin M~D 2011 {\em Phys.
  Rev. Lett.\/} {\bf 107} 133602
  \urlprefix\url{http://link.aps.org/doi/10.1103/PhysRevLett.107.133602}

\bibitem{Chen_2013}
Chen W, Beck K~M, B{\"u}cker R, Gullans M, Lukin M~D, Tanji-Suzuki H and
  Vuleti{\'c} V 2013 {\em Science\/} {\bf 341} 768--770
  \urlprefix\url{http://www.sciencemag.org/content/341/6147/768.abstract}

\bibitem{friedler_deterministic_2005}
Friedler I, Kurizki G and Petrosyan D 2005 {\em Phys. Rev. A\/} {\bf 71} 023803
  \urlprefix\url{http://link.aps.org/doi/10.1103/PhysRevA.71.023803}

\bibitem{ParedesBarato_AllOptical_2014}
Paredes-Barato D and Adams C~S 2014 {\em Phys. Rev. Lett.\/} {\bf 112}(4)
  040501 \urlprefix\url{http://link.aps.org/doi/10.1103/PhysRevLett.112.040501}

\bibitem{firstenberg_attractive_2013}
Firstenberg O, Peyronel T, Liang Q~Y, Gorshkov A~V, Lukin M~D and Vuleti{\'c} V
  2013 {\em Nature\/} {\bf 502} 71--75 ISSN 0028-0836
  \urlprefix\url{http://www.nature.com/nature/journal/v502/n7469/full/nature12512.html}

\bibitem{walker_consequences_2008}
Walker T~G and Saffman M 2008 {\em Phys. Rev. A\/} {\bf 77} 032723
  \urlprefix\url{http://link.aps.org/doi/10.1103/PhysRevA.77.032723}

\bibitem{nipper_highly_2012}
Nipper J, Balewski J~B, Krupp A~T, Butscher B, L{\"o}w R and Pfau T 2012 {\em
  Phys. Rev. Lett.\/} {\bf 108} 113001
  \urlprefix\url{http://link.aps.org/doi/10.1103/PhysRevLett.108.113001}

\bibitem{ryabtsev_forster_2010}
Ryabtsev I~I, Tretyakov D~B, Beterov I~I and Entin V~M 2010 {\em Phys. Rev.
  Lett.\/} {\bf 104}(7) 073003
  \urlprefix\url{http://link.aps.org/doi/10.1103/PhysRevLett.104.073003}

\bibitem{anderson_resonant_1998}
Anderson W~R, Veale J~R and Gallagher T~F 1998 {\em Phys. Rev. Lett.\/} {\bf
  80} 249 \urlprefix\url{http://link.aps.org/doi/10.1103/PhysRevLett.80.249}

\bibitem{Ditzhuijzen08}
van Ditzhuijzen C~S~E, Koenderink A~F, Hern\'andez J~V, Robicheaux F, Noordam
  L~D and van~den Heuvell H~B~v~L 2008 {\em Phys. Rev. Lett.\/} {\bf 100}(24)
  243201 \urlprefix\url{http://link.aps.org/doi/10.1103/PhysRevLett.100.243201}

\bibitem{mudrich_transfer_2010}
Mudrich M, Zahzam N, Vogt T, Comparat D and Pillet P 2005 {\em Phys. Rev.
  Lett.\/} {\bf 95}(23) 233002
  \urlprefix\url{http://link.aps.org/doi/10.1103/PhysRevLett.95.233002}

\bibitem{gunter_observing_2013}
G\"{u}nter G, Schempp H, Robert-de Saint-Vincent M, Gavryusev V, Helmrich S,
  Hofmann C~S, Whitlock S and Weidem\"{u}ller M 2013 {\em Science\/} {\bf 342}
  954--956 ISSN 0036-8075, 1095-9203
  \urlprefix\url{http://www.sciencemag.org/content/342/6161/954}

\bibitem{ravets_2014}
Ravets S, Labuhn H, Barredo D, B{\'e}guin L, Lahaye T and Browaeys A 2014 {\em
  arXiv preprint arXiv:1405.7804\/}

\bibitem{Entangled_EIT_2014}
Li W, Viscor D, Hofferberth S and Lesanovsky I 2014 {\em Phys. Rev. Lett.\/}
  {\bf 112}(24) 243601
  \urlprefix\url{http://link.aps.org/doi/10.1103/PhysRevLett.112.243601}

\bibitem{dudin_emergence_2012}
Dudin Y~O, Bariani F and Kuzmich A 2012 {\em Phys. Rev. Lett.\/} {\bf 109}
  133602 \urlprefix\url{http://link.aps.org/doi/10.1103/PhysRevLett.109.133602}

\bibitem{dudin_strongly_2012}
Dudin Y~O and Kuzmich A 2012 {\em Science\/} {\bf 336} 887–--889

\bibitem{wilk_entanglement_2010}
Wilk T, Gaëtan A, Evellin C, Wolters J, Miroshnychenko Y, Grangier P and
  Browaeys A 2010 {\em Phys. Rev. Lett.\/} {\bf 104} 010502
  \urlprefix\url{http://link.aps.org/doi/10.1103/PhysRevLett.104.010502}

\bibitem{Isenhower_CNOT_2010}
Isenhower L, Urban E, Zhang X~L, Gill A~T, Henage T, Johnson T~A, Walker T~G
  and Saffman M 2010 {\em Phys. Rev. Lett.\/} {\bf 104}(1) 010503
  \urlprefix\url{http://link.aps.org/doi/10.1103/PhysRevLett.104.010503}

\bibitem{Wu_EIT_VdWinteracion_2014}
Wu H, Bian M~M, Shen L~T, Chen R~X, Yang Z~B and Zheng S~B 2014 {\em Phys. Rev.
  A\/} {\bf 90}(4) 045801
  \urlprefix\url{http://link.aps.org/doi/10.1103/PhysRevA.90.045801}

\bibitem{Carroll_angular_2004}
Carroll T~J, Claringbould K, Goodsell A, Lim M~J and Noel M~W 2004 {\em Phys.
  Rev. Lett.\/} {\bf 93}(15) 153001
  \urlprefix\url{http://link.aps.org/doi/10.1103/PhysRevLett.93.153001}

\bibitem{reinhard_level_2007}
Reinhard A, Liebisch T~C, Knuffman B and Raithel G 2007 {\em Phys. Rev. A\/}
  {\bf 75} 032712
  \urlprefix\url{http://link.aps.org/doi/10.1103/PhysRevA.75.032712}

\bibitem{saffman_efficient_2009}
Saffman M and M\o{}lmer K 2009 {\em Phys. Rev. Lett.\/} {\bf 102}(24) 240502
  \urlprefix\url{http://link.aps.org/doi/10.1103/PhysRevLett.102.240502}

\bibitem{Barredo_3atom_anisotropic_2014}
Barredo D, Ravets S, Labuhn H, B\'eguin L, Vernier A, Nogrette F, Lahaye T and
  Browaeys A 2014 {\em Phys. Rev. Lett.\/} {\bf 112}(18) 183002
  \urlprefix\url{http://link.aps.org/doi/10.1103/PhysRevLett.112.183002}

\bibitem{Linington_ComplexDetuning_2008}
Linington I~E and Vitanov N~V 2008 {\em Phys. Rev. A\/} {\bf 77}(6) 062327
  \urlprefix\url{http://link.aps.org/doi/10.1103/PhysRevA.77.062327}

\bibitem{Moiseev_QuantumMemory_2004}
Moiseev S~A and Ham B~S 2004 {\em Phys. Rev. A\/} {\bf 70}(6) 063809
  \urlprefix\url{http://link.aps.org/doi/10.1103/PhysRevA.70.063809}

\bibitem{fleischhauer_dark-state_2000}
Fleischhauer M 2000 {\em Phys. Rev. Lett.\/} {\bf 84} 5094–5097

\bibitem{scully_quantum_1997}
Scully M~O and Zubairy M~S 1997 {\em Quantum Optics\/} 1st ed (Cambridge
  University Press) ISBN 0521435951

\bibitem{sevincli_nonlocal_2011}
Sevin\ifmmode~\mbox{\c{c}}\else \c{c}\fi{}li S, Henkel N, Ates C and Pohl T
  2011 {\em Phys. Rev. Lett.\/} {\bf 107}(15) 153001
  \urlprefix\url{http://link.aps.org/doi/10.1103/PhysRevLett.107.153001}

\bibitem{Harris_NormalModes_1994}
Harris S~E 1994 {\em Phys. Rev. Lett.\/} {\bf 72}(1) 52--55
  \urlprefix\url{http://link.aps.org/doi/10.1103/PhysRevLett.72.52}

\bibitem{dudin_observation_2012}
Dudin Y~O, Li L, Bariani F and Kuzmich A 2012 {\em Nat. Phys.\/} {\bf 8}
  790--794 \urlprefix\url{http://dx.doi.org/10.1038/nphys2413}

\bibitem{maxwell_storage_2013}
Maxwell D, Szwer D~J, Paredes-Barato D, Busche H, Pritchard J~D, Gauguet A,
  Weatherill K~J, Jones M~P~A and Adams C~S 2013 {\em Phys. Rev. Lett.\/} {\bf
  110} 103001
  \urlprefix\url{http://link.aps.org/doi/10.1103/PhysRevLett.110.103001}

\bibitem{Konopnicki_Simultons_1981}
Konopnicki M~J and Eberly J~H 1981 {\em Phys. Rev. A\/} {\bf 24}(5) 2567--2583
  \urlprefix\url{http://link.aps.org/doi/10.1103/PhysRevA.24.2567}

\bibitem{Harris_MatchedEIT_1993}
Harris S~E 1993 {\em Phys. Rev. Lett.\/} {\bf 70}(5) 552--555
  \urlprefix\url{http://link.aps.org/doi/10.1103/PhysRevLett.70.552}

\bibitem{Eberly_DressedField_1994}
Eberly J~H, Pons M~L and Haq H~R 1994 {\em Phys. Rev. Lett.\/} {\bf 72}(1)
  56--59 \urlprefix\url{http://link.aps.org/doi/10.1103/PhysRevLett.72.56}

\bibitem{Grobe_Adiabatons_1994}
Grobe R, Hioe F~T and Eberly J~H 1994 {\em Phys. Rev. Lett.\/} {\bf 73}(24)
  3183--3186
  \urlprefix\url{http://link.aps.org/doi/10.1103/PhysRevLett.73.3183}

\bibitem{Fleischhauer_Pulsematching_1995}
Fleischhauer M and Richter T 1995 {\em Phys. Rev. A\/} {\bf 51}(3) 2430--2442
  \urlprefix\url{http://link.aps.org/doi/10.1103/PhysRevA.51.2430}

\bibitem{Xiao-xue_PropagatingEntangledState_2005}
xue Yang X and Wu Y 2005 {\em Journal of Optics B: Quantum and Semiclassical
  Optics\/} {\bf 7} 54
  \urlprefix\url{http://stacks.iop.org/1464-4266/7/i=2/a=004}

\bibitem{Viscor_Phaseonium_2011}
Viscor D, Ferraro A, Loiko Y, Mompart J and Ahufinger V 2011 {\em Phys. Rev.
  A\/} {\bf 84}(4) 042314
  \urlprefix\url{http://link.aps.org/doi/10.1103/PhysRevA.84.042314}

\bibitem{Scully_phaseonium_1992}
Scully M~O 1992 {\em Physics Reports\/} {\bf 219} 191 -- 201 ISSN 0370-1573
  \urlprefix\url{http://www.sciencedirect.com/science/article/pii/037015739290136N}

\bibitem{Choi_Mapping_2008}
Choi K~S, Deng H, Laurat J and Kimble H~J 2008 {\em Nature\/} {\bf 452} 67--71

\end{thebibliography}

\end{document}